\documentstyle{mn}

\input psfig.sty

\title[Cooling in self-gravitating protoplanetary discs]
	{The effect of cooling on the global stability of \\ 
	self-gravitating protoplanetary discs}

\author[Rice et al.]{W.K.M. Rice$^1$, P.J. Armitage$^{2,3}$, M.R. Bate$^4$ and I.A. Bonnell$^1$ \\
	$^1$School of Physics and Astronomy, University 
	of St Andrews, North Haugh, St Andrews KY16 9SS \\
	$^2$JILA, Campus Box 440, University of Colorado, Boulder CO 80309-0440, USA \\
	$^3$Department of Astrophysical and Planetary Sciences, University of Colorado, Boulder CO 80309-0391, USA \\ 
	$^4$School of Physics, University of Exeter, Stocker Road, Exeter EX4 4QL}		
\begin{document}

\maketitle

\begin{abstract}
Using a local model Gammie (2001) has shown that accretion discs with cooling times 
$t_{\rm cool} \le 3 \Omega^{-1}$ fragment into gravitationally bound objects, while
those with cooling times $t_{\rm cool} > 3 \Omega^{-1}$ evolve into a quasi-steady state. We
use three-dimensional smoothed particle hydrodynamic simulations of protoplanetary
accretion discs to test if the local results hold globally. We find that for disc masses
appropriate for T Tauri discs, the fragmentation boundary still occurs at a cooling time
close to $t_{\rm cool} = 3 \Omega^{-1}$. For more massive discs, which are likely to be
present at an earlier stage of the star formation process, fragmentation occurs for
longer cooling times, but still within a factor of two of that predicted using a local model. 
These results have implications not only for planet formation in protoplanetary discs
and star formation in AGN discs, but also for the redistribution of angular momentum which
could be driven by the presence of relatively massive objects within the accretion disc.
\end{abstract}


\section{Introduction}
The formation of substellar companions within protoplanetary discs has received
a great deal of attention since the first observation of an extra-solar planet 
(Mayor \& Queloz 1995). To date many additional extrasolar planets have been observed
(see Marcy \& Butler 2000 for a review), further fueling the interest in the formation of
planets and the evolution of protoplanetary discs. The most widely studied 
planet formation mechanism is core accretion 
(see Lissauer 1993) in which planetismals grow by direct collisions to form a core which, 
when sufficiently massive ($m \sim 10 m_{\rm earth}$) then accretes an envelope of
gas from the disc. 

An alternative mechanism for giant planet formation is via the
gravitational instability (Kuiper 1951; Boss 1998, 2000). This differs from core
accretion in that a rocky core is not initially required and the process is
extremely rapid. A Keplerian accretion disc 
with sound speed $c_s$, surface density $\Sigma$, 
and epicyclic frequency $\kappa$ will become gravitationally unstable if the 
Toomre (1964) $Q$ parameter
\begin{equation}
Q=\frac{c_s \kappa}{\pi G \Sigma}
\end{equation}
is of order unity.  A gravitationally unstable disc can either fragment into one or 
more gravitationally bound objects, or it can evolve into a quasi-stable state
in which gravitational instabilities lead to the outward transport of angular 
momentum. The exact outcome depends on the rate at which the disc heats up (through
the dissipation of turbulence and gravitational instabilities) 
and the rate at which the disc cools. It has been
suggested (Goldreich \& Lynden-Bell 1965) that a feedback loop may exist where when
$Q$ is large cooling dominates and the disc is cooled towards instability. When
$Q$ becomes sufficiently small, heating through viscous dissipation dominates and the
disc is returned to a state of marginal stability. In this way $Q$ is maintained at a value
of $\sim 1$.

Gammie (2001) has, however, shown using a local model that a quasi-stable state can only
be maintained if the cooling time $t_{cool} > 3 \Omega^{-1}$ where $\Omega$ is the 
local angular frequency.  For shorter cooling times, the disc fragments. This finding is
consistent with Pickett et al. (1998, 2000) that `almost isothermal' conditions are
necessary for fragmentation, and defines a robust lower limit to the critical cooling time
below which fragmentation occurs. It has been suggested, however, that self-gravitating
discs strictly require a global treatment (Balbus \& Papaloizou 1999), and while global effects
are highly unlikely to stabilize a locally unstable disc, they could well allow fragmentation
within discs that would locally be stable.  
In this paper we use a global model to test whether a gaseous accretion disc still fragments
for $t_{cool} \le 3 \Omega^{-1}$ and attains a quasi-stable state for $t_{cool} > 3 \Omega^{-1}$.  
Although our choice of parameters is more appropriate for protoplanetary
accretion discs, the results may also be applicable for star formation in AGN discs, 
which are expected to become gravitationally unstable at large radii (Shlosman \& Begelman
1989; Goodman 2002), and for star formation in the gaseous regions of spiral galaxies
(Kim \& Ostriker 2002).  

\section{Numerical simulations}

\subsection{Smooth particle hydrodynamics code}
The three-dimensional simulations presented here were performed using smoothed 
particle hydrodynamics (SPH), a Lagrangian hydrodynamics code 
(e.g., Benz 1990; Monaghan 1992).  In this simulation the central 
star is modelled as a point mass onto which gas particles can accrete if they
approach to within the sink radius, while the gaseous disc is simulated using
250,000 SPH particles. Both point masses and gas use a tree to
determine neighbours and to calculate gravitational forces (Benz et al. 1990), and
the point mass representing the central star 
is free to move under the gravitational influence of the disc gas.
Although the code can in principle continue running if the disc starts to fragment 
and high density regions form, it quickly becomes too slow to realistically continue
in this manner.  To continue simulating a fragmenting disc, point masses are created 
(Bate et al. 1995) if the high density
regions are gravitationally bound 
(gravitational potential energy at least twice the thermal energy). As with the
point mass representing the central star, point masses within the disc may
continue to accrete gas particles if they fall within the sink radius. 
An additional saving in computational time is also made by using individual,
particle time-steps (Bate et al. 1995; Navarro \& White 1993).  The time-steps
for each particle are limited by the Courant condition and a force condition
(Monaghan 1992).  

\subsection{Initial conditions}
We consider a system comprising a central star, modelled as a point mass with mass $M_*$, 
surrounded by a gaseous
circumstellar disc with mass $M_{\rm disc}$. Most of the simulations presented here
considered a disc mass of $M_{\rm disc} = 0.1 M_*$, but a few were performed using 
$M_{\rm disc} = 0.25 M_*$.  The disc temperature is taken to
have an initial radial profile of $T \propto r^{-0.5}$ (e.g., Yorke \& Bodenheimer 1999) 
and the Toomre $Q$ parameter is
assumed to be initially constant with a value of $2$. A stable accretion disc where 
self-gravity leads to the
steady outward transportation of angular momentum should have a near constant Q
throughout.  A constant $Q$ together with equation (1) then
gives a surface density profile of $\Sigma \propto r^{-7/4}$, and hydrostatic
equilibrium in the vertical direction gives a central density profile of 
$\rho \propto r^{-3}$.  

The disc is modelled using 250,000 SPH
particles, which are initially randomly distributed in such a way as to
give the specified 
density profile between inner and outer radii of $r_{\rm in}$ and 
$r_{\rm out}$ respectively. 

The calculations performed here are essentially scale free. In code units, 
we take $M_* = 1$, $r_{\rm in} = 1$ and $r_{\rm out} =25$. If we were to assume
a physical mass scale of $1 M_{\odot}$  and a length scale of $1$ au, the central
star would have a mass of $1 M_{\odot}$, 
the circumstellar disc would have a mass of $0.1 M_{\odot}$, or $0.25 M_{\odot}$, 
and would extend from 
1~au to 25~au, and 1 year would equal $2 \pi$ code units.

\subsection{Cooling}
Since the aim of this work is to test whether the results obtained
by Gammie (2001) using a local model still hold globally, we use the same
approach in our global model as used in the local model. 
We use an adiabatic equation of state, with adiabatic index $\gamma = 5/3$, and 
allow the disc gas to heat up due to both PdV work and viscous 
dissipation. Cooling is implemented by adding a simple cooling 
term to the energy equation. Specifically, for a particle with 
internal energy per unit mass $u_i$,
\begin{equation} 
 { {{\rm d} u_i} \over {{\rm d}t} } = - {u_i \over {t_{\rm cool}}}
\end{equation}  
where, as in Gammie (2001), $t_{\rm cool}$ is given by $\beta \Omega^{-1}$ 
with the value of $\beta$ varied for each run.  

Although the use of the above cooling time is essentially chosen to compare
the local model results with results using a global model, it can also
be related (at least approximately) to the real physics of an 
accretion disc.
For an optically thick disc in equilibrium, the cooling
time is the ratio of the thermal energy per unit area to the radiative losses
per unit area. It can be
shown (e.g., Pringle 1981) that in such a viscous accretion disc, 
the cooling time is given by 
\begin{equation}
 t_{\rm cool} = \frac{4}{9\gamma(\gamma-1)}\frac{1}{\alpha \Omega}
\label{tcool}
\end{equation}
where $\gamma$ is the adiabatic index, $\Omega$ is the angular frequency, and
$\alpha$ is the Shakura \& Sunyaev (1973) viscosity parameter. 

\section{Results}

\subsection{$M_{\rm disk} = 0.1 M_*$}
\begin{figure}
\centerline{\psfig{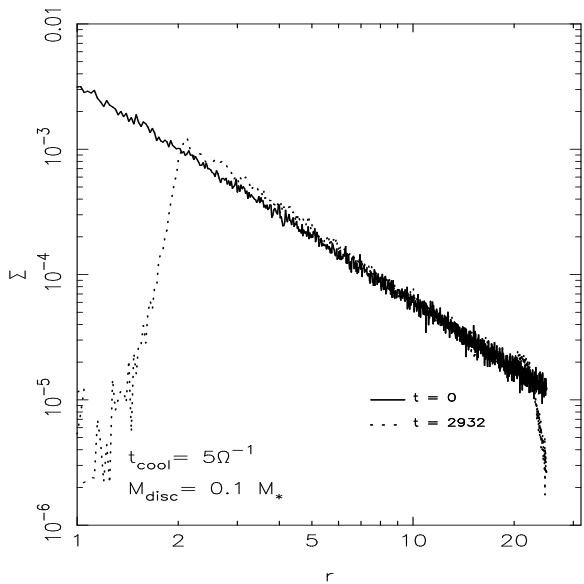}}
\caption{\label{surfdens}Surface density for $t_{cool}=5 \Omega^{-1}$ and 
$M_{\rm disc} = 0.1 M_*$ at the
beginning and end of the simulation.  Apart from particles with initial radii
between 1 and 2 being accreted onto the central star, the surface density does
not change significantly during the simulation.}
\end{figure}

We use a global model to consider how cooling times of  
$t_{\rm cool} = 5 \Omega^{-1}$, and $t_{\rm cool} = 3 \Omega^{-1}$ affect the 
gravitational stability of a viscous accretion disc with a mass of $M_{\rm disc} = 0.1 M_*$. 
Figure
\ref{surfdens} shows the disc surface density, $\Sigma$, at the beginning and end of the
$t_{\rm cool} = 5 \Omega^{-1}$ simulation. Since we have not attempted to model the inner
boundary condition in any detail, there is rapid accretion and a drop in surface density
close to the inner boundary. Apart from particles with radii between 1 and 2 being accreted
onto the central star, the surface density profile does not change significantly during the
course of the simulation. This also occurs for $t_{\rm cool} = 3
\Omega^{-1}$ and a corresponding figure is consequently not shown.

\begin{figure}
\centerline{\psfig{figure=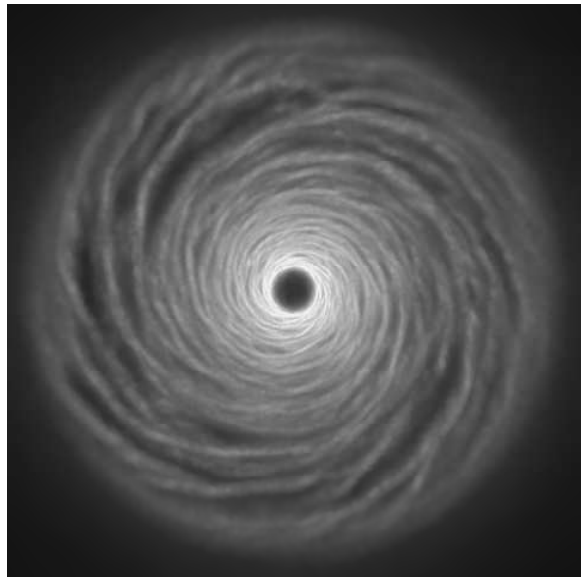,width=3.0truein,height=3.0truein}}
\caption{\label{tc5disc}Equatorial density structure for $t_{cool} = 5 \Omega^{-1}$ 
and $M_{\rm disc} = 0.1 M_*$. 
The disc is highly structure with the instability existing at all radii. The
density has, however, not increased significantly and the disc is in a quasi-stable state 
with
heating through viscous dissipation balancing cooling.}
\end{figure}

\begin{figure}
\centerline{\psfig{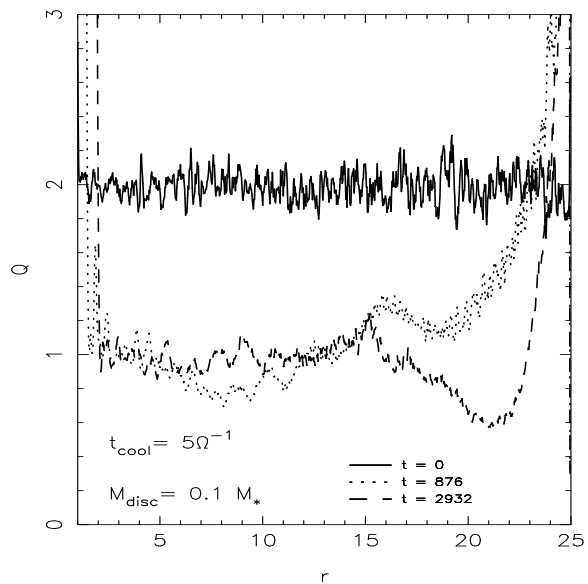}}
\caption{\label{tc5Q}Toomre $Q$ parameter at the beginning ($t = 0$), one third of the way
($t = 876$), and at the end ($t = 2932$) of the $t_{\rm cool} = 5 \Omega^{-1}$,
$M_{\rm disc} = 0.1 M_*$ simulation. At the
end of the simulation $Q \sim 1$ at radii between 1 and 15.  A low $Q$ region also moves
to larger radii with time, indicating the current radius of maximum instability growth.}
\end{figure}

Figure \ref{tc5disc} shows the final equatorial density structure of the 
$t_{\rm cool} = 5 \Omega^{-1}$ 
simulation. The central star (not shown) is located in the middle of the figure 
and the $x$ and $y$
axes both run from -25 to 25. Figure \ref{tc5Q} shows the Toomre $Q$ parameter for
the same simulation at the beggining ($t=0$), approximately a third of the way into
the simulation ($t = 876$), and at the end ($t = 2932$) of the simulation. After $t = 2932$
time units, particles at the outer edge of the disk ($r = 25$) have completed 3.7 
orbits while particles at $r = 1$ have completed 466 orbits. The disc is highly structured
and the instability exists at all radii. However, at no point in the 
disc has the density increased significantly and no fragmentation has taken place.  Figure
\ref{tc5Q} shows that at $t = 2932$ (dashed line) the Toomre $Q$ parameter, which initially
had a constant value of 2 (solid line), is close to 1 for radii between 1 and 15. 
Comparing $Q$ at $t = 876$ and at $t = 2932$ suggests that the cooling may initially reduce Q
to below 1, causing the gravitational instability to grow, heating the disc and returning
$Q$ to a value of order unity. Figure \ref{tc5Q} also shows this region of low $Q$ moving
to larger radii with time. This is a consequence of both the cooling time and the dynamical
time (the timescale on which heating occurs) both depending directly on radius. The 
instability therefore starts at the inner radii and by the end of the simulation has 
reached the edge ($r = 25$) of the disc. The disc has reached a quasi-steady state in
which cooling is balanced by heating through viscous dissipation resulting from the growth
of the gravitational instability.

\begin{figure}
\centerline{\psfig{figure=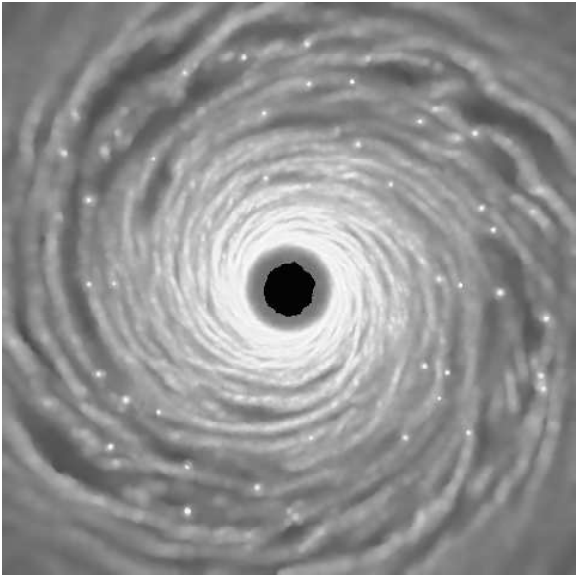,width=3.0truein,height=3.0truein}}
\caption{\label{tc3disc}Equatorial density structure for $t_{cool} = 3 \Omega^{-1}$ and
$M_{\rm disc} = 0.1 M_*$. The
disc is highly unstable and is fragmenting. The fragments are all gravitationally bound.}
\end{figure}

\begin{figure}
\centerline{\psfig{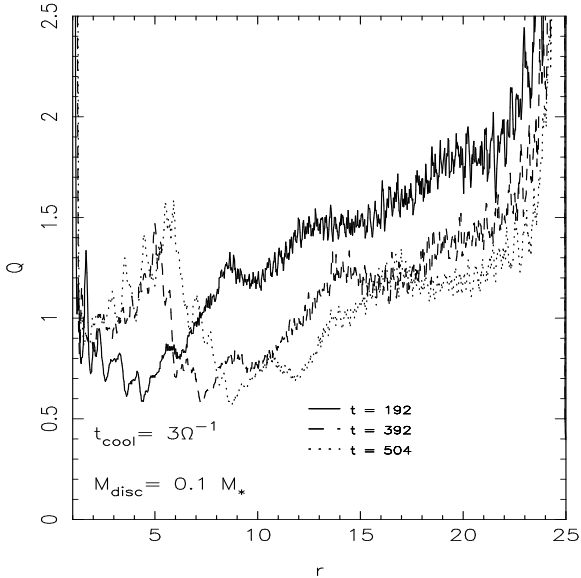}}
\caption{\label{tc3Q}Toomre $Q$ parameter at times of $t = 102$, 
$t = 362$, and $t = 504$ for $t_{\rm cool} = 3 \Omega^{-1}$ and $M_{\rm disc} = 0.1 M_*$. 
The value of $Q$ is reduced such that fragmentation of the
disc takes place. A region of low $Q$ moves
to larger radii with time, indicating the current fragmentation radius.}
\end{figure}

Figure \ref{tc3disc} shows the final equatorial density structure of the $t_{\rm cool} = 
3 \Omega^{-1}$ simulation. We were only able to run this simulation for a total of
$504$ time units and hence Figure \ref{tc3disc} shows only the inner 8 radii of the simulation.
The central star is again in the middle of the figure. Figure \ref{tc3Q} shows the Toomre
$Q$ parameter at three different times during the simulation. The initial $Q$ for
$t_{\rm cool} = 3 \Omega^{-1}$ is the same as for 
$t_{\rm cool} = 5 \Omega^{-1}$ and is therefore not shown in Figure \ref{tc3Q}. 
Figure \ref{tc3disc}
shows that not only is the disc highly structured, the bright dots also indicate that 
fragmentation is taking place.  The
fragments that can be seen in Figure \ref{tc3disc} are all gravitationally bound. 
To reach the time shown in Figure \ref{tc3disc} it was necessary to 
convert most of the fragments into point masses (Bate et al. 1995). 
Figure \ref{tc3Q} shows that at $t = 192$ $Q$ is less than 1 for radii
between 1 and 5. At later times the minimum $Q$ value moves to larger radii and the 
value of $Q$ at smaller radii increases to a value of between 1 and 1.5. As in the
$t_{\rm cool} = 5 \Omega^{-1}$ simulation, the gravitational instability grows from the 
inner radii and moves to larger radii with time. While $Q$ is below
1, the instability grows rapidly, producing gravitationally bound fragments. 
These fragments tidally interact with the disc, heating it and causing $Q$ to increase 
to a value greater than unity. Once $Q > 1$ fragmentation ceases, and
the disc, at that radius, becomes gravitationally stable. There is, however, likely to
be a significant amount of angular momentum transport driven by tidal interactions between
fragments and the gaseous disc (Larson 2002). 

Simulations using $t_{\rm cool} = 10 \Omega^{-1}$ and 
$t_{\rm cool} = 4 \Omega^{-1}$ were also performed.
Both simulations were run for more than 2000 time units, almost $4$ times longer than the
$t_{\rm cool} = 3 \Omega^{-1}$ simulation.  The $t_{\rm cool} =10 \Omega^{-1}$
simulation showed no noticeable structure and the Toomre $Q$ settled down to a value
between $1$ and $2$. For $t_{\rm cool} = 4 \Omega^{-1}$ the disc structure was similar to
that obtained using $t_{\rm cool} = 5 \Omega^{-1}$ except for the presence of a few high density 
clumps at radii of $\sim 10$ and greater. By the end of the simulation ($t = 2136$) one 
of the clumps, at a radius of $9.5$,
had satisfied the condition for point mass creation, i.e. it was gravitationally bound. This
would suggest that, globally, fragmentation may occur at slightly greater cooling time than 
expected using a local model (Gammie 2001). However, unlike the $t_{\rm cool} = 3 \Omega^{-1}$
simulation, 
the $t_{\rm cool} = 4 \Omega^{-1}$ simulation 
did not fragment at all radii, with the inner $\sim 10$ radii of the disc remaining in
a quasi-stable state.  
 
\subsection{$M_{\rm disc} = 0.25 M_*$}

A simulation using a heavier disc ($M_{\rm disc} = 0.25 M_*$) was also performed.  Although
most T Tauri discs have masses significantly less than this (Beckwith et al. 1990), 
there are a few with comparable masses. Such a simulation may also be appropriate
for an earlier stage of the star formation process when the discs are expected to be heavier. 
We considered cooling times of $t_{cool} = 10 \Omega^{-1}$
and $t_{\rm cool} = 5 \Omega^{-1}$. The $t_{\rm cool} = 10 \Omega^{-1}$ simulation
was run for $1024$ time units,
at the end of which, as for $M_{\rm disc} = 0.1 M_*$, the disc showed minimal structure and there 
was no sign of any 
density enhancement or fragmentation. 

\begin{figure}
\centerline{\psfig{figure=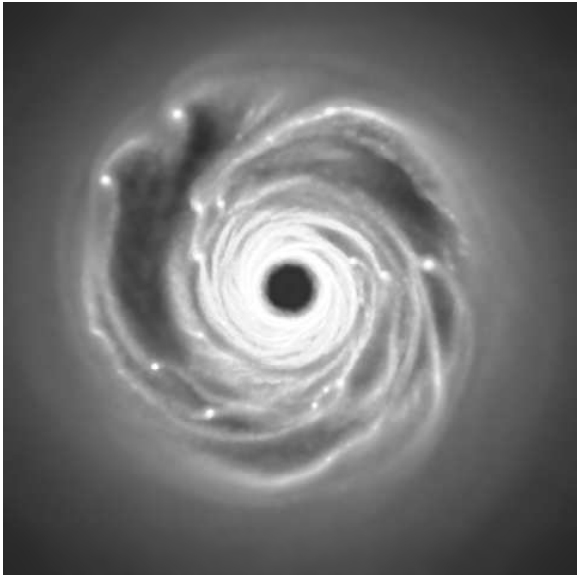,width=3.0truein,height=3.0truein}}
\caption{\label{tc5Md025disc}Equatorial density structure for $t_{cool} = 5 \Omega^{-1}$ and
for a disc mass $M_{\rm disc} = 0.25 M_*$. There are
signs of fragmentation with the most massive fragment being gravitationally bound.}
\end{figure}

The $t_{\rm cool} = 5 \Omega^{-1}$ simulation was run for $684$ time units and the 
final equatorial density structure is shown in Figure \ref{tc5Md025disc}. The disc 
is highly structured and the spirals,
consistent with the increased disc mass (Nelson et al. 1998), are somewhat less
filamentary then for the $M_{\rm disc} = 0.1 M_*$ simulation with the same cooling time. Unlike
the equivalent $M_{\rm disc} = 0.1 M_*$ simulation (see Figure \ref{tc5disc}) there are, however, 
a number of high density clumps present in the disc. The routine for converting these clumps
into point masses checks the nearest $\sim 50$ SPH particles to determine if they are 
gravitationally bound (see Bate et al. 1995). Using this technique, the high density regions in
Figure \ref{tc5Md025disc} were found to be gravitationally unbound and could not be 
converted into point masses. The simulation was eventually stopped at $t = 684$ when
the maximum density had increased to a value $10^7$ times greater than the initial maximum
density. 
Restricting the determination of the boundness of the clump to only the nearest $\sim 50$ 
SPH particles is somewhat arbitrary. By including the nearest $\sim 250$ SPH particles, rather
than only the nearest $\sim 50$, the densest clump was found to be just gravitationally bound,
i.e. the gravitational potential energy was almost exactly twice the thermal energy.  This
would imply that in more massive discs, global effects could act to make the disc more 
unstable, allowing gravitationally bound fragments to grow for cooling times greater than 
that obtained using
a local model. This may have implications both for the formation of planets, or
binary companions (Adams, Ruden \& Shu 1989), reasonably
early in the star formation process.

\subsection{Cooling time and effective viscosity}
\begin{figure}
\centerline{\psfig{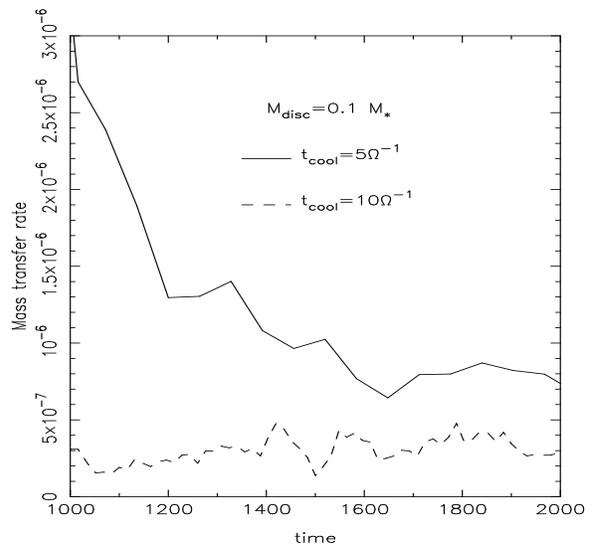}}
\caption{\label{masstrans}A running average of the magnitude of the 
mass transfer rate for $t_{\rm cool} = 5 \Omega^{-1}$ and $t_{\rm cool} = 10 \Omega^{-1}$
and for a disc mass $M_{\rm disc} = 0.1 M_*$.}
\end{figure}

By using a radially dependent cooling time in our simulations, Equation (\ref{tcool})
implies that the discs should settle into a state in which the effective viscous $\alpha$
(Shakura \& Sunyaev 1973) depends inversely on this cooling time. Since none of our
simulations were run to a steady state, which would take several viscous times to
achieve, it is not straightforward to determine whether the local scaling of $\alpha$
with $t_{\rm cool}$ is obeyed. This is further complicated in the $t_{\rm cool} =
3 \Omega^{-1}$, $M_{\rm disc} = 0.1 M_*$ simulation since the collapsed mass
fraction increases throughout the run. We are, however able to compare the 
$t_{\rm cool} = 
5 \Omega^{-1}$, $M_{\rm disc} = 0.1 M_*$ simulation 
with the $t_{\rm cool} = 10 \Omega^{-1}$,
$M_{\rm disc} = 0.1 M_*$ simulation. Figure \ref{masstrans} shows the magnitude of 
the mass transfer rate through a radius $r = 15$, plotted against time in code units, for 
$t_{\rm cool} = 5 \Omega^{-1}$ (solid line) and $t_{\rm cool} = 10 \Omega^{-1}$ 
(dashed line)
and a disc mass, in both cases, of $M_{\rm disc} = 0.1 M_*$. Figure \ref{masstrans}
only considers the mass transfer rate once the instability has almost saturated at 
the radius considered. Since neither simulation has reached a steady state and
since both discs are self-gravitating, the mass transfer rate can be negative or 
positive. Figure \ref{masstrans} is consequently a $60$ time unit running average of the
magnitude of the mass transfer rate and shows that the 
mass transfer rate for $t_{\rm cool} = 
5 \Omega^{-1}$ is, as expected, higher than that for $t_{\rm cool} = 10 \Omega^{-1}$.
By the end of simulation the mass transfer rates are both reasonably constant and
differ by a factor of between $2$ and $3$. Although this difference is 
consistent with that expected from Equation (\ref{tcool}), the complications in
trying to accurately determine the relationship between $\alpha$ and $t_{\rm cool}$
in these global simulations leads us to simply conclude that, as expected, the
effective viscosity does indeed increase with decreasing cooling time.  

\section{Conclusion}
Using a local model, Gammie (2001) has shown that for cooling times $t_{\rm cool} \le 3 
\Omega^{-1}$ a disc will fragment into one or more gravitationally bound objects, while for
longer cooling times the disc will settle into a quasi-stable state with heating through
viscous dissipation balancing cooling. We have performed three-dimensional disc
calculations using SPH to test if the above still holds globally. We use the same
cooling function as used by Gammie (2001) which, although fairly simplistic, can also
be justified physically (Pringle 1981).   
 
For a disc mass of $M_{\rm disc} = 0.1 M_*$ and for 
cooling times of $t_{\rm cool} = 5 \Omega^{-1}$ and
$t_{\rm cool} = 10 \Omega^{-1}$ we find that the disc settles into a quasi-stable state. At the
end of the $t_{\rm cool} = 5 \Omega^{-1}$ run, the Toomre $Q$ parameter is of order unity 
between radii of 1
and 15 and the disc is highly structured.  The density is, however, not significantly greater
than the initial density and no fragmentation has occured. A comparison between the
$t_{\rm cool} = 5 \Omega^{-1}$ and the $t_{\rm cool} = 
10 \Omega^{-1}$ simulations also
shows that once the instability has saturated at a given radius, 
the magnitude of the mass transfer rate was greatest for the smaller cooling
time. Although we were unable
to quantify the relationship between the cooling time and the viscous $\alpha$ (Shakura
\& Sunyaev 1973), this result shows that, as expected, the effective viscosity does
indeed increase with decreasing cooling time.

For $t_{\rm cool} = 3 \Omega^{-1}$
the disc rapidly becomes unstable and produces numerous gravitationally bound
fragments. The growth of these fragments starts near the inner 
boundary of the disc and had we been able to continue with the simulation, it seems
likely that fragmentation would have continued throughout the disc. A 
$t_{\rm cool} = 4 \Omega^{-1}$ simulation also produced gravitationally bound fragments but these
were located at radii of $\sim 10$ or greater. The inner radii of the disc remained in
a quasi-steady state. 

A simulation using a heavier disc ($M_{\rm disc} = 0.25 M_*$) 
was found to fragment for cooling
times of $t_{\rm cool} = 5 \Omega^{-1}$, suggesting that as the disc mass 
increases, global effects may act to make the disc more unstable. A cooling time
of $t_{\rm cool} = 10 \Omega^{-1}$, however, showed no signs of fragmentation and hence
the fragmentation boundary, even for the heavier disc, is still within a factor of $\sim 2$ 
of that obtained using a local model (Gammie 2001). For T Tauri discs, which generally
have masses $M_{\rm disc} < 0.1 M_*$ (Beckwith et  al. 1990), the fragmentation
boundary is therefore likely to be close to that determined using the local model. 
Earlier in the star
formation process, when the discs are expected to be more massive, 
fragmentation may occur for 
cooling times somewhat longer than that predicted by the local model results (Gammie 2001). 

For quiescent T Tauri discs, $\alpha$ is conventionally estimated to be of the order of
$10^{-2}$ or smaller (Hartmann et al. 1998; Bell \& Lin 1994). The simple estimates quoted
earlier would suggest that such discs are comfortably stable. Boss (2001), however, has shown 
using an approximate treatment of the disc heating and cooling that there may be periods when
the cooling time is comparable to the orbital period. Our results suggest that if the disc
is fairly massive, such short cooling times open a window of opportunity for the formation
of substellar objects, probably in the form of a multiple system (Armitage \& Hansen 1999). 
Not only does this have implications for
planet formation in protoplanetary discs, it also has implications for angular momentum 
transport. The presence of a number of massive clumps within a protoplanetary disc is
likely to significantly enhance the transport of angular momentum through tidal interactions
(Larson 2002). Similar considerations apply to AGN discs. Numerous theoretical studies
(e.g. Clarke 1988; Kumar 1999; Hure 2000; Menou \& Quataert 2001) show that AGN discs
invariably become self-gravitating at radii of $\sim 0.1 \ {\rm pc}$. Fragmentation would lead
to the formation of stars (Collin \& Zahn 1999), and truncation of the inner accretion disc
(Goodman 2002). Instability of gaseous discs at much larger radii of $10-1000$ pc may also
occur (Shlosman \& Begelman 1989). Fragmentation of these larger scale discs would yield
a flattened stellar system, while rapid angular momentum transport could play a role in
replenishing the inner galactic resevoir (Shlosman \& Begelman 1989).

\section*{Acknowlegments}

The simulations reported in this paper made use of the 
UK Astrophysical Fluids Facility (UKAFF). WKMR acknowledges 
support from a PPARC Standard Grant.

\end{document}